\documentclass[aps,pre,preprint,superscriptaddress]{revtex4}
\usepackage{graphics,epsfig,wasysym}

\begin{document}

\centerline{\large\bf Density Functional Theory of a Curved Liquid-Vapour Interface:}
\centerline{\large\bf Evaluation of the rigidity constants}
\vskip 20pt
\centerline{Edgar M. Blokhuis$^1$ and A.E. van Giessen$^2$}
\vskip 5pt
\centerline{\it $^1$Colloid and Interface Science, Leiden Institute of Chemistry,}
\centerline{\it Gorlaeus Laboratories, P.O. Box 9502, 2300 RA Leiden, The Netherlands}
\vskip 5pt
\centerline{\it $^2$Hobart \& William Smith College, Department of Chemistry, Geneva, NY 14456, USA}
\vskip 15pt
\centerline{\bf Abstract}
\vskip 5pt
\noindent
It is argued that to arrive at a quantitative description
of the surface tension of a liquid drop as a function of its inverse 
radius, it is necessary to include the bending rigidity $k$ and
Gaussian rigidity $\bar{k}$ in its description. New formulas for
$k$ and $\bar{k}$ in the context of density functional theory with
a non-local, integral expression for the interaction between molecules
are presented. These expressions are used to investigate the influence
of the choice of Gibbs dividing surface and it is shown that for a
one-component system, the {\em equimolar} surface has a special status in
the sense that both $k$ and $\bar{k}$ are then the {\em least} sensitive to a
change in the location of the dividing surface. Furthermore, the equimolar
value for $k$ corresponds to its {\em maximum} value and the equimolar
value for $\bar{k}$ corresponds to its {\em minimum} value. An explicit
evaluation using a short-ranged interaction potential between molecules,
shows that $k$ is {\em negative} with a value around minus 0.5-1.0 $k_{\rm B} T$
and that $\bar{k}$ is {\em positive} with a value which is a bit
more than half the magnitude of $k$. Finally, for dispersion forces between
molecules, we show that a term proportional to $\log(R) / R^2$
replaces the rigidity constants and we determine the (universal)
proportionality constants.
\vskip 5pt
\noindent
\centerline{\rule{300pt}{1pt}}

\section{Introduction}

\noindent
The surface tension of a simple drop of liquid has captured the imagination of
scientists dating back to the pioneering work of J. Williard Gibbs \cite{Gibbs}.
This interest continues with the main focus of attention directed
towards the description of the deviation of the surface tension from
its planar value when the radius of the liquid droplet becomes smaller.
Such a deviation is especially important in the theoretical description
of nucleation phenomena \cite{nucleation}. The homogeneous nucleation of a
liquid from a supersaturated vapour follows via the formation of small
liquid droplets and the nucleation time and energy depend sensitively
on the precise value of the droplet's surface tension.

A key quantity in quantifying the extent by which the surface tension of
a liquid drop deviates from its planar value is the {\em Tolman length}
introduced by Tolman in 1949 \cite{Tolman}. It can be defined in two equivalent
ways. In the first way, one considers the radial dependence of the surface
tension of a (spherical) liquid droplet defined as the excess grand free
energy per unit area:
\begin{equation}
\Omega = -p_{\ell} \, V_{\ell} - p_v \, V_v + \sigma_s(R) \, A \,.
\end{equation}
When the radius $R$ of the droplet is large, the surface tension may be expanded
in the inverse radius:
\begin{equation}
\label{eq:sigma_s}
\sigma_s(R) = \sigma - \frac{2 \delta \sigma}{R} + \ldots \,, 
\end{equation}
where $\sigma$ is the surface tension of the {\em planar interface} and
where the leading order correction defines the Tolman length $\delta$.
In the second route to define the Tolman length, one considers the pressure difference
$\Delta p \!=\! p_{\ell} - p_v$ between the pressure of the liquid inside and
the pressure of the vapour outside the droplet. For large radii of curvature,
$\Delta p$ is expanded in $1/R$:
\begin{equation}
\label{eq:Delta_p}
\Delta p = \frac{2 \sigma}{R} - \frac{2 \delta \sigma}{R^2} + \ldots \,. 
\end{equation}
The first term on the right hand side is the familiar Laplace equation \cite{RW}
with the leading order correction giving Tolman's original definition of the
Tolman length \cite{Tolman}. It is important to note that this correction only
takes on the form in Eq.(\ref{eq:Delta_p}) when the {\em equimolar} radius
\cite{Gibbs} is taken as the radius of the liquid drop, i.e. $R \!=\! R_e$.
Furthermore, with this choice of the (Gibbs) dividing surface, terms of
order ${\cal O}(1/R^3)$ are absent and the dots represent terms of order
${\cal O}(1/R^4)$. When the location of the droplet radius is chosen {\em away}
from the equimolar radius, the Tolman length correction to the Laplace equation
has a form different than that shown in Eq.(\ref{eq:Delta_p}). For instance,
the radius corresponding to the so-called {\em surface of tension}
($R \!=\! R_s$) is defined such that Eq.(\ref{eq:Delta_p}) appears
as $\Delta p \!=\! 2 \sigma(R_s) / R_s$.

The determination of the value of the Tolman length for a simple drop of
liquid has proved to be not without controversy (recent reviews are given in
refs. \cite{Blokhuis06, Malijevsky12}). This is mainly due to two reasons: first,
one of the first microscopic expressions for the Tolman length was formulated
in the context of a {\em mechanical approach} which lead to an expression
for the Tolman length in terms of the first moment of the excess tangential
pressure profile of a planar interface \cite{Buff}. However, it was pointed out
by Henderson and Schofield in 1982 that such an expression depends on the form 
of the pressure tensor used and is therefore not well-defined \cite{Hemingway81,
Henderson82, Schofield82, Henderson_book}. Furthermore, even the evaluation of the Tolman
length using the usual Irving-Kirkwood \cite{IK} form for the pressure tensor
leads to {\em incorrect} results \cite{Blokhuis92b} and the use of the mechanical
expression is now (mostly) abandoned.

A second origin of controversy is simply due to the fact that for a regular liquid-vapour
interface the Tolman length is {\em small} (a fraction of the molecular diameter),
since it measures the subtle asymmetry between the liquid and vapour phase.
Straightforward squared-gradient theory with the familiar $tanh$-profile for
the density profile, leads to a {\em zero} value of the Tolman length \cite{FW, Blokhuis93}
and it remains a challenge to distinguish its value from zero in computer simulations
\cite{Nijmeijer, Frenkel, Bardouni00, Lei, Horsch12}. Nowadays, those computer simulations
that have succeeded in obtaining a value different from zero indicate that its value
is {\em negative} with its magnitude around one tenth of a molecular diameter
\cite{Giessen09, Binder09, Sampoyo10, Binder10, Binder11, Binder12} and error
bars usually somewhat less than half that number.

The sign and magnitude of the Tolman length for a regular liquid-vapour interface are
corroborated by a large number of different versions of density functional theory (DFT),
which has proved to be an invaluable tool in the theoretical description of inhomogeneous
systems \cite{Sullivan, Evans79, Evans84, Evans90}. Quite surprisingly, the details of
the density functional theory at hand do not seem to matter that much \cite{Bykov06,
Malijevsky12} and one ubiquitously finds that the Tolman length is {\em negative} with
a magnitude comparable to that obtained in simulations.
This includes results for the Tolman length from van der Waals squared-gradient theory
\cite{Baidakov99, Baidakov04a}, density functional theory with a non-local, integral
expression for the interaction between molecules (DFT-LDA) \cite{Malijevsky12,
Giessen98, Koga98, Napari01, Barrett06}, density functional theory with
weighted densities (DFT-WDA) \cite{Bykov06} and density functional theory using Rosenfeld's
\cite{Rosenfeld} fundamental measure theory for the hard-sphere free energy (DFT-FMT)
\cite{Li08, Sampoyo10, Binder10, Binder12}.

All in all, there now seems to be the same level of agreement between simulations and
DFT for the Tolman length as it exists for the surface tension, with the exception of
one particular type of simulation result. In refs. \cite{Giessen09, Binder09, Sampoyo10,
Binder10, Binder11, Binder12} the Tolman length is determined in computer simulations
of liquid droplets for various (large) radii of curvature, but in a different
set of simulations the Tolman length is extracted from computer simulations of a
{\em planar interface} \cite{Haye, Giessen02}, using a virial expression
for the Tolman length \cite{Blokhuis92a}. The simulations of the planar interface
lead to a Tolman length that has the same order of magnitude as the simulations of
the liquid droplets but now with the {\em opposite} sign. It has been suggested
that, since the interfacial area is much larger in the simulations of the planar
interface, the presence of {\em capillary waves} might play an important role
\cite{Giessen09}. However, it is difficult to imagine that this would change
the sign of the Tolman length so that the resolution to this problem remains
uncertain.

Another feature that ubiquitously results from the computer simulations and DFT
calculations of liquid droplets is that the surface tension is {\em not monotonous}
as a function of the (inverse) radius (for a recent review, see ref.~\cite{Malijevsky12}).
A {\em maximum} in the surface tension of a liquid droplet occurs which suggests
that the surface tension is qualitatively better approximated by a {\em parabola}
rather than by a straight line with its slope given by the Tolman length.
This means that one needs to include higher order terms, going beyond the level
of the Tolman length, in the expansion of the surface tension in Eq.(\ref{eq:sigma_s}).
Such an expansion was first provided in the ground-breaking work by Helfrich
in 1973 \cite{Helfrich}. The form for the free energy suggested by Helfrich
is the most general form for the surface free energy of an isotropic surface
expanded to second order in the surface's curvature \cite{Helfrich}:
\begin{equation}
\label{eq:Helfrich}
\Omega_{\rm H} = \int \!\! dA \; [ \, \sigma - \delta \sigma \, J
+ \frac{k}{2} \, J^2 + \bar{k} \, K \, + \ldots ] \,,
\end{equation}
where $J \!=\! 1/R_1 + 1/R_2$ is the total curvature, $K \!=\! 1/(R_1 R_2)$
is the Gaussian curvature and $R_1$, $R_2$ are the principal radii of curvature
at a certain point on the surface. The expansion defines four curvature coefficients:
$\sigma$, the surface tension of the planar interface, $\delta$, the Tolman
length \cite{Tolman}, $k$, the bending rigidity, and $\bar{k}$, the rigidity
constant associated with Gaussian curvature. The original expression proposed
by Helfrich \cite{Helfrich} features the radius of spontaneous curvature $R_0$ as the
linear curvature term ($\delta \sigma \rightarrow 2 k / R_0$ \cite{Blokhuis06, Blokhuis92b}),
but in honour of Tolman we stick to the notation in Eq.(\ref{eq:Helfrich}).

For surfaces for which the curvatures $J$ and $K$ are constant, the Helfrich
free energy per unit area reduces to:
\begin{equation}
\label{eq:sigma(J,K)}
\Omega_{\rm H}/A \equiv \sigma(J,K) = \sigma - \delta \sigma \, J
+ \frac{k}{2} \, J^2 + \bar{k} \, K + \ldots \,,
\end{equation}
which for a spherically or cylindrically shaped surface takes the form:
\begin{eqnarray}
\label{eq:sigma_s(R)}
\sigma_s(R) &=& \sigma - \frac{2 \delta \sigma}{R} 
+ \frac{(2 k + \bar{k})}{R^2} + \ldots \hspace*{27pt} {\rm (sphere)} \\
\label{eq:sigma_c(R)}
\sigma_c(R) &=& \sigma - \frac{\delta \sigma}{R} 
+ \frac{k}{2 R^2} + \ldots \hspace*{56pt} {\rm (cylinder)}
\end{eqnarray}
These expressions indicate that the second order coefficients, which express
the non-monotonicity of the surface tension as observed in simulations and DFT
calculations of liquid drops, are given by the combination of the rigidity
constants $2 k + \bar{k}$ and the bending rigidity $k$. Our goal in this
article is to provide general formulas for the bending rigidities $k$ and
$\bar{k}$ using density functional theory (DFT-LDA). This work extends
previous work by us \cite{Giessen98}, by Koga and Zeng \cite{Koga99},
by Barrett \cite{Barrett09} and by Baidakov {\em et al.} \cite{Baidakov04b}.
Our formulas are  subsequently applied to explicitly evaluate the bending
rigidities and it is determined how well they can be used to describe the
surface tension of a liquid drop (or vapour bubble).

The expansion of the surface tension of a liquid drop to second order in
$1/R$ has not been without controversy \cite{Henderson92, Rowlinson94, Fisher}.
Two issues have played a role here. The first issue concerns the fact that when
the interaction between molecules is sufficiently long-ranged, the expansion
in $1/R$ may not be analytic beyond some term \cite{Blokhuis92a, Hooper, Dietrich}.
In particular, for {\em dispersion forces} the second order correction has
the form $\log(R) / R^2$ rather than $1/R^2$ and one could argue that the
rigidity constants are ``infinite''. Nowadays, this point is well-appreciated
and no longer source of controversy. In this article we come back to this
issue and provide explicit expressions for the second order correction to
replace the expansion in Eq.(\ref{eq:sigma_s(R)}) or (\ref{eq:sigma_c(R)})
for dispersion forces.

A second issue argues that {\em even for short-ranged interactions}, which
are mostly considered in simulations and DFT calculations, the second order
term might pick up a logarithmic correction of the form $\log(R) / R^2$
\cite{Henderson92, Rowlinson94, Fisher}.
The reasoning behind this focuses on the fact that for a {\em spherical}
droplet, the second order contribution to the free energy, i.e. the expression
in Eq.(\ref{eq:sigma_s(R)}) multiplied by the area $A \!=\! 4 \pi \, R^2$
is {\em independent} of $R$, which might be an indication that it should be
replaced by a logarithmic term. The most compelling argument {\em against}
this reasoning lies in the fact that the same argument applied to a {\em cylindrical}
interface would lead to the conclusion that already the linear term in $1/R$
(Tolman length) would pick up logarithmic corrections. Although the issue
is not completely settled, the presence of a logarithmic correction for
short-ranged interaction has not been observed in simulations or
demonstrated in calculations either in mean-field theory (DFT) or
in Statistical Mechanics \cite{Blokhuis92a}. Also in this article, we
inspect (numerically) the possible presence of a logarithmic correction
to the second order term in the expansion of the free energy of a liquid
drop and find no evidence for its presence.

Our article is organized as follows: in the next section we discuss the
density functional theory that is considered (DFT-LDA) and use it to
determine the surface tension $\sigma_s(R)$ of a liquid drop and vapour bubble.
In Section \ref{sec-expansion}, the free energy is expanded to second
order in $1/R$ for a spherical and cylindrical interface which allows the
formulation of new, closed expressions for the rigidity constants $k$ and
$\bar{k}$ \cite{Giessen98, Barrett09}. An important feature addressed is
the consequence of the choice made for the location of the dividing surface
(the value of $R$) on the value of the bending rigidities. The formulas
for $k$ and $\bar{k}$ are explicitly evaluated using a cut-off and shifted
Lennard-Jones potential for the attractive part of the interaction potential.
Since the evaluation of these expressions requires numerical determination
of the density profile, we supply in Section \ref{sec-explicit} an accurate
approximation based on squared-gradient theory to evaluate $\delta$, $k$
and $\bar{k}$ from the parameters of the phase diagram only.
In Section \ref{sec-dispersion} we consider the full Lennard-Jones
interaction potential and determine its consequences for the expansion
of the free energy in $1/R$. We end with a discussion of results.

\section{Density functional theory}
\label{sec-DFT}

\noindent
The expression for the (grand) free energy in density functional theory
is based on the division into a hard-sphere reference system plus
attractive forces described by an interaction potential $U_{\rm att}(r)$.
It is the following functional of the density $\vec{r}$ \cite{Sullivan,
Evans79, Evans84, Evans90}:
\begin{equation}
\label{eq:Omega_DFT}
\Omega[\rho] = \int \!\! d\vec{r} \; [ \; f_{\rm hs}(\rho) - \mu \rho(\vec{r}) \; ]
+ \frac{1}{2} \int \!\! d\vec{r}_1 \! \int \!\! d\vec{r}_{12} \;
U_{\rm att}(r) \, \rho(\vec{r}_1) \rho(\vec{r}_2) \,,
\end{equation} 
where $\mu$ is the chemical potential. For the free energy of the
hard-sphere reference system $f_{\rm hs}(\rho)$, we take the well-known
Carnahan-Starling form \cite{CS}:
\begin{equation}
\label{eq:g_hs}
f_{\rm hs}(\rho) = k_{\rm B} T \, \rho \, \ln(\rho) 
+ k_{\rm B} T \, \rho \, \frac{(4 \eta - 3 \eta^2)}{(1 - \eta)^2} \,,
\end{equation} 
where $\eta \!\equiv\! (\pi/6) \, \rho \, d^3$ with $d$ the
molecular diameter. The Euler-Lagrange equation that minimizes
the free energy in Eq.(\ref{eq:Omega_DFT}) is given by:
\begin{equation}
\label{eq:EL_DFT}
\mu = f^{\prime}_{\rm hs}(\rho) + \int \!\! d\vec{r}_{12} \; U_{\rm att}(r) \, \rho(\vec{r}_2) \,.
\end{equation}
For a {\em uniform} system, the Euler-Lagrange equation becomes:
\begin{equation}
\label{eq:mu}
\mu = f^{\prime}_{\rm hs}(\rho) - 2 a \, \rho \,,
\end{equation} 
with the van der Waals parameter $a$ explicitly expressed in terms
of the interaction potential as
\begin{equation}
\label{eq:a}
a \equiv - \frac{1}{2} \int \!\! d\vec{r}_{12} \; U_{\rm att}(r) \,.
\end{equation}
Using the expression for the chemical potential in Eq.(\ref{eq:mu}),
the bulk pressure is obtained from $\Omega \!=\! - p V$ leading to
the following equation of state:
\begin{equation}
\label{eq:EOS}
p = \frac{k_{\rm B} T \, \rho \, (1 + \eta + \eta^2 - \eta^3)}{(1-\eta)^3} - a \, \rho^2 \,.
\end{equation} 
Next, we consider the implementation of DFT in planar and spherical geometry.
\vskip 10pt
\noindent
{\bf Planar interface}
\vskip 5pt
\noindent
When the chemical potential is chosen such that a liquid and vapour phase coexist,
$\mu \!=\! \mu_{\rm coex}$, a planar interface forms between the two phases. The
density profile is then a function of the coordinate normal to the interface,
$\rho(\vec{r}) \!=\! \rho_0(z)$. In planar geometry, the Euler-Lagrange equation in
Eq.(\ref{eq:EL_DFT}) becomes:
\begin{equation}
\label{eq:EL_planar}
\mu_{\rm coex} = f^{\prime}_{\rm hs}(\rho_0) + \int \!\! d\vec{r}_{12} \; U_{\rm att}(r) \, \rho_0(z_2) \,.
\end{equation} 
The surface tension of the planar interface is the surface free energy
per unit area ($\sigma \!=\! (\Omega + p \, V)/A$ \cite{RW}):
\begin{equation}
\label{eq:sigma_DFT}
\sigma = - \frac{1}{4} \int\limits_{-\infty}^{\infty} \!\!\! dz_1 \! \int \!\! d\vec{r}_{12} \;
U_{\rm att}(r) \, r^2 (1-s^2) \, \rho_0^{\prime}(z_1) \rho_0^{\prime}(z_2) \,,
\end{equation}
where $z_2 \!=\! z_1 + sr$ and $s \!=\! \cos \theta_{12}$.
\vskip 10pt
\noindent
{\bf A Spherical Drop of Liquid}
\vskip 5pt
\noindent
When the chemical potential $\mu$ is varied to a value {\em off-coexistence},
spherically shaped liquid droplets in {\em metastable} equilibrium with a
bulk vapour phase may form. Such droplets are termed {\em critical} droplets.
The radius of the liquid droplet is taken to be equal to the {\em equimolar} radius,
$R \!=\! R_e$ \cite{Gibbs}, which depends on the value of the chemical potential
chosen, and is defined as:
\begin{equation}
\label{eq:R}
4 \pi \int\limits_{0}^{\infty} \!\! dr \; r^2 \left[ \, \rho_s(r) - \rho_v \right]
= \frac{4 \pi}{3} \, R_e^3 \, (\rho_{\ell} - \rho_v) \,.
\end{equation}
The (grand) free energy for the formation of the critical droplet is given by:
\begin{equation}
\label{eq:Delta_Omega_0}
\frac{\Delta \Omega}{A} \equiv \frac{\Omega + p_v \, V}{A} = - \frac{\Delta p \, R}{3} + \sigma_s(R) \,,
\end{equation}
with $p_v$ the vapour pressure outside the droplet and $p_{\ell} = p_v + \Delta p$
is the liquid pressure inside (see the remark below, however). The surface tension
of the critical droplet is the quantity that we wish to study and this equation provides
a way to determine it from $\Delta \Omega$.

In spherical geometry, the free energy density functional in Eq.(\ref{eq:Omega_DFT}) is given by:
\begin{eqnarray}
\label{eq:Delta_Omega}
\frac{\Delta \Omega[\rho_s]}{A} &=& \int\limits_{0}^{\infty} \!\! dr_1 \left( \frac{r_1}{R} \right)^{\!2}
[ \; f_{\rm hs}(\rho_s) - \mu \rho_s(r_1) \; ] \\
&& + \frac{1}{2} \int\limits_{0}^{\infty} \!\! dr_1 \left( \frac{r_1}{R} \right)^{\!2} \! \int \!\! d\vec{r}_{12} \;
U_{\rm att}(r) \, \rho_s(r_1) \rho_s(r_2) \,, \nonumber
\end{eqnarray}
with the Euler-Lagrange equation that minimizes the above free energy equal to:
\begin{equation}
\label{eq:EL_sphere}
\mu = f^{\prime}_{\rm hs}(\rho_s) + \int \!\! d\vec{r}_{12} \; U_{\rm att}(r) \, \rho_s(r_2) \,.
\end{equation}
The procedure to determine $\sigma_s(R)$ as a function of $R$ is as follows:
\vskip 5pt
\noindent
{\bf (1)} First, the bulk densities $\rho_{0,\ell}$ and $\rho_{0,v}$ and the
chemical potential at two-phase coexistence, $\mu_{\rm coex}$, are determined
by solving the following set of equations: 
\begin{equation}
\label{eq:bulk_0}
f^{\prime}(\rho_{0,v}) = \mu_{\rm coex} \,, \hspace*{10pt}
f^{\prime}(\rho_{0,\ell}) = \mu_{\rm coex} \,, \hspace*{10pt}
f(\rho_{0,v}) - \mu_{\rm coex} \, \rho_{0,v} = f(\rho_{0,\ell}) - \mu_{\rm coex} \, \rho_{0,\ell} \,,
\end{equation}
where we have defined $f(\rho) \!\equiv\! f_{\rm hs}(\rho) - a \rho^2$. The bulk density
difference is denoted as $\Delta \rho \!\equiv\! \rho_{0,\ell} - \rho_{0,v}$ and the
pressure at coexistence is simply $p_{\rm coex} \!=\! -f(\rho_{0,\ell/v}) +
\mu_{\rm coex} \, \rho_{0,\ell/v}$.
\vskip 5pt
\noindent
{\bf (2)} Next, the chemical potential $\mu$ is varied to a value {\em off-coexistence}.
For $\mu \!>\! \mu_{\rm coex}$ liquid droplets are formed ($R \!>\! 0$)
and when $\mu \!<\! \mu_{\rm coex}$ we obtain bubbles of vapour ($R \!<\! 0$).
For given temperature and chemical potential $\mu$ the liquid and vapour
densities $\rho_{\ell}$ and $\rho_v$ are then determined from solving the
following two equations
\begin{equation}
f^{\prime}(\rho_{v}) = \mu \,, \hspace*{25pt} f^{\prime}(\rho_{\ell}) = \mu \,,
\end{equation}
with the corresponding bulk pressures calculated from
\begin{equation}
\label{eq:pressures}
p_{v} = - f(\rho_{v}) + \mu \, \rho_{v} \,, \hspace*{25pt}
p_{\ell} = - f(\rho_{\ell}) + \mu \, \rho_{\ell} \,.
\end{equation}
It should be remarked that far outside the droplet ($r \!\rightarrow\! \infty$),
the density (or pressure) is equal to that of the bulk,
$\rho_s(\infty) \!=\! \rho_{v}$, but that only for large droplets
is the density {\em inside} the droplet ($\rho_s(r \!=\! 0)$) equal to
its bulk value ($\rho_{\ell}$).
\vskip 5pt
\noindent
{\bf (3)} Finally, the Euler-Lagrange equation for $\rho_s(r)$ in
Eq.(\ref{eq:EL_sphere}) is solved numerically with the boundary condition
$\rho_s(\infty) \!=\! \rho_{v}$. The resulting density profile
$\rho_s(r)$ is inserted into Eq.(\ref{eq:R}) to determine the equimolar
radius $R \!=\! R_e$ and into Eq.(\ref{eq:Delta_Omega}) to determine
$\Delta \Omega$ and thus $\sigma_s(R)$.

\begin{figure}
\centering
\includegraphics[angle=270,width=250pt]{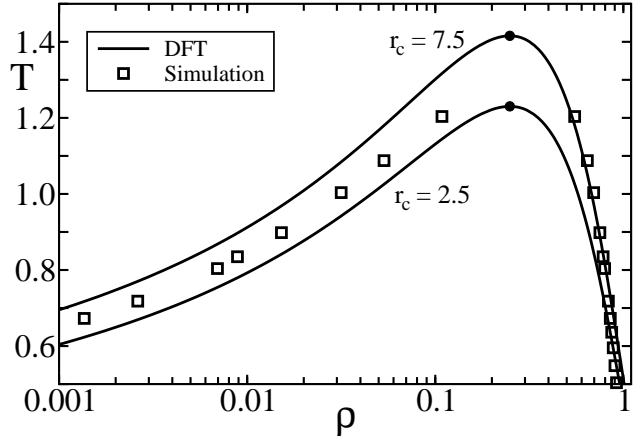}
\caption{Phase diagram as a function of reduced temperature and density.
The solid lines are the liquid-vapour densities at two values of the reduced
LJ cut-off radius (solid circles indicate the location of the critical points).
Square symbols are simulation results from ref.~\cite{Baidakov07}.}
\label{Fig:pd}
\end{figure}

\vskip 10pt
\noindent
This procedure is carried out using a cut-off and shifted Lennard-Jones
potential for the attractive part of the interaction potential:
\begin{eqnarray}
\label{eq:LJ}
U_{\rm att}(r) =  \left\{
\begin{array}{cc}
U_{\rm LJ}(r_{\rm min}) - U_{\rm LJ}(r_c) & \hspace*{100pt} 0 < r < r_{\rm min} \\
U_{\rm LJ}(r) - U_{\rm LJ}(r_c)           & \hspace*{100pt} r_{\rm min} < r < r_c \\
0                                         & \hspace*{100pt} r > r_c 
\end{array}
\right.
\end{eqnarray}
where $U_{\rm LJ}(r) \!=\! 4 \varepsilon \, [ \, (d/r)^{12} - (d/r)^6 \, ]$ and
$r_{\rm min} \!=\! 2^{\frac{1}{6}} \, d$. Figure \ref{Fig:pd} shows the resulting phase
diagram as a function of reduced density $\rho^* \!\equiv\! \rho \, d^3$ and reduced
temperature $T^* \!\equiv\! k_{\rm B} T / \varepsilon$. The solid lines are the liquid-vapour
densities for two values of the LJ cut-off radius; the square symbols are recent
computer simulation results taken from ref.~\cite{Baidakov07}.

\begin{figure}
\centering
\includegraphics[angle=270,width=250pt]{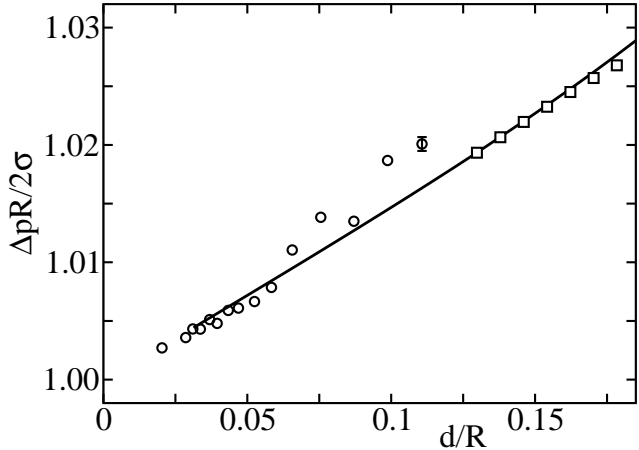}
\caption{Pressure difference multiplied by $R / 2 \sigma$ as a function
of the reciprocal {\em equimolar} radius $d/R$. Circular symbols are simulation
results from ref.~\cite{Giessen09}. DFT calculations are shown as the solid
line ($\Delta p \!=\! p(0) - p_v$) and square symbols ($\Delta p \!=\! p_{\ell} - p_v$).
For the DFT calculations we have set the reduced temperature $T^* \!=\!$ 0.911297
and reduced LJ cut-off $r_c \!=$ 2.5. The value for the reduced temperature is
chosen such that the liquid-vapour density difference at coexistence matches
the value in the computer simulations \cite{Giessen09}.}
\label{Fig:Delta_p}
\end{figure}

In Figure \ref{Fig:Delta_p}, we show the pressure difference multiplied
by $R / 2 \sigma$ as a function of the reciprocal radius. The circular
symbols are previous simulation results \cite{Giessen09} that were
used to determine the Tolman length from (minus) the slope at
$1/R \!=\! 0$ ($\delta \!\approx$ - 0.10 $d$ \cite{Giessen09}).
For comparison, we show the result of DFT calculations as the solid
line, where we have taken the pressure at the center of the droplet
as the liquid pressure. The excellent agreement in Figure \ref{Fig:Delta_p}
is somewhat misleading since the corresponding values of the surface
tension differ by as much as 50 \%. As square symbols, the results of
DFT calculations using $p_{\ell}$ from Eq.(\ref{eq:pressures})
as the liquid pressure are plotted to show that the slight difference
between $p(0)$ and $p_{\ell}$ for small droplets has no consequences
for the determination of $\delta$. 

\begin{figure}
\centering
\includegraphics[angle=270,width=250pt]{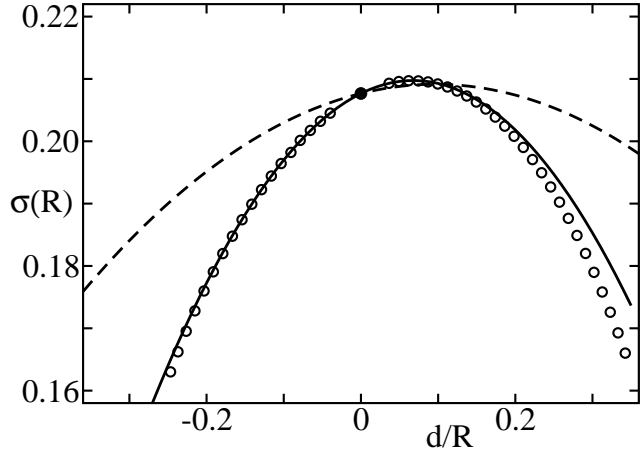}
\caption{Droplet surface tension (in units of $k_{\rm B} T / d^2$) as a function
of the reciprocal {\em equimolar} radius $d/R$; vapour bubbles are formed for
$R \!<\! 0$ and liquid droplets for $R \!>\! 0$. The solid line is the
parabolic approximation to $\sigma_s(R)$ determined from the expansion in Section
\ref{sec-expansion}. As a comparison, the parabolic approximation to the cylindrical
surface tension $\sigma_c(R)$ is shown as the dashed line. We have set the reduced
temperature $T^* \!=\!$ 1.0 and reduced LJ cut-off $r_c \!=$ 2.5.}
\label{Fig:sigma_R}
\end{figure}

In Figure \ref{Fig:sigma_R}, a typical example of the surface tension of a spherical
liquid drop (and vapour bubble) is shown as a function of $1/R$, with $R$ the
equimolar radius of the droplet. The symbols are the values for $\sigma_s(R)$
calculated using DFT. The solid line is the parabolic approximation in 
Eq.(\ref{eq:sigma_s(R)}) with values for the coefficients $\sigma$, $\delta$,
and $2 k + \bar{k}$ calculated from formulas presented in the next Section. 
The behaviour of the surface tension is characterized by a positive first
derivative at $1/R \!=\!0$, which indicates that the Tolman length is {\em negative},
and a negative {\em second} derivative which indicates that the combination
$2 k + \bar{k}$ is also {\em negative}. It is concluded that the parabolic
approximation gives a {\em quantitatively} accurate description for the surface
tension for a large range of reciprocal radii. The determination of the full
$\sigma_s(R)$ is usually quite elaborate and it therefore seems sufficient
to only determine the coefficients in the parabolic approximation to $\sigma_s(R)$
as a function of $1/R$. This is done in the next Section.

\section{Curvature expansion}
\label{sec-expansion}

\noindent
In this section, we consider spherically and cylindrically shaped liquid droplets
and expand the free energy and density profile systematically to second order
in $1/R$. An important feature of our analysis will be to not restrict ourselves
to a particular choice of the dividing surface, but to instead leave the radius
$R$ unspecified. This will allow us to derive new, more general expressions and
will allow for a new investigation of the consequences of varying the choice for
the location of the dividing surface.

To second order in $1/R$, the expansion of the density profile of the spherical
droplet reads:
\begin{equation}
\label{eq:expansion_rho}
\rho_s(r) =  \rho_0(z) + \frac{1}{R} \, \rho_{s,1}(z) + \frac{1}{R^2} \, \rho_{s,2}(z) + \ldots \,,
\end{equation}
where $z \!=\! r - R$. The leading order correction to the density
profile of the spherical interface is twice that of the cylindrical
interface, so it is convenient to define
$\rho_1(z) \!\equiv\! \rho_{s,1}(z) \!=\! 2 \, \rho_{c,1}(z)$.
We shall consider the expansion of the free energy of the spherical
and cylindrical droplet separately.
\vskip 10pt
\noindent
{\bf Spherical interface}
\vskip 5pt
\noindent
The coefficients in the curvature expansion of the density are determined
from the curvature expansion of the Euler-Lagrange equation in
Eq.(\ref{eq:EL_sphere}). The result is that the (planar) density
profile $\rho_0(z)$ is determined from Eq.(\ref{eq:EL_planar}) and $\rho_1(z)$
follows from solving:
\begin{equation}
\label{eq:EL_1}
\mu_1 = f^{\prime\prime}_{\rm hs}(\rho_0) \, \rho_1(z_1) + \int \!\! d\vec{r}_{12} \; U_{\rm att}(r) \,
[ \, \rho_1(z_2) + \frac{r^2}{2} (1-s^2) \, \rho^{\prime}_0(z_2) \, ] \,,
\end{equation} 
where $\mu_1 \!=\! 2 \sigma / \Delta \rho$ \cite{Blokhuis93, Blokhuis06}.
For the evaluation of the curvature coefficients it turns out to be sufficient
to determine the density profiles $\rho_0(z)$ and $\rho_1(z)$ only.

The expansion for $\rho_s(r)$ is inserted into the expression for the
free energy in Eq.(\ref{eq:Delta_Omega}). Performing a systematic expansion
to second order in $1/R$, using the Euler-Lagrange equations in
Eqs.(\ref{eq:EL_planar}) and (\ref{eq:EL_1}), one ultimately
obtains expressions for the curvature coefficients by comparing the free energy
to the curvature expansion in Eq.(\ref{eq:sigma_s(R)}). For the surface tension of
the planar interface the result in Eq.(\ref{eq:sigma_DFT}) is recovered:
\begin{equation}
\label{eq:sigma}
\sigma = - \frac{1}{4} \int\limits_{-\infty}^{\infty} \!\!\! dz_1 \! \int \!\! d\vec{r}_{12} \;
U_{\rm att}(r) \, r^2 (1-s^2) \, \rho_0^{\prime}(z_1) \rho_0^{\prime}(z_2) \,.
\end{equation} 
For the Tolman length one obtains the following expression \cite{Giessen98}
\begin{equation}
\label{eq:delta}
\delta \sigma = \frac{1}{4} \int\limits_{-\infty}^{\infty} \!\!\! dz_1 \! \int \!\! d\vec{r}_{12} \;
U_{\rm att}(r) \, r^2 (1-s^2) \, z_1 \, \rho_0^{\prime}(z_1) \rho_0^{\prime}(z_2)
- \frac{\mu_1}{2} \int\limits_{-\infty}^{\infty} \!\!\! dz \; z \, \rho_0^{\prime}(z) \,.
\end{equation}
For the combination of the rigidity constants, $2k + \bar{k}$, we have:
\begin{eqnarray}
\label{eq:k_sph}
2k + \bar{k} &=& \frac{1}{4} \int\limits_{-\infty}^{\infty} \!\!\! dz_1 \! \int \!\! d\vec{r}_{12} \;
U_{\rm att}(r) \, r^2 (1-s^2) \, \rho_0^{\prime}(z_1) \rho_1(z_2) \\
&-& \frac{1}{4} \int\limits_{-\infty}^{\infty} \!\!\! dz_1 \! \int \!\! d\vec{r}_{12} \;
U_{\rm att}(r) \, r^2 (1-s^2) \, z_1^2 \, \rho_0^{\prime}(z_1) \rho_0^{\prime}(z_2) \nonumber \\
&+& \frac{1}{48} \int\limits_{-\infty}^{\infty} \!\!\! dz_1 \! \int \!\! d\vec{r}_{12} \;
U_{\rm att}(r) \, r^4 (1-s^4) \, \rho_0^{\prime}(z_1) \rho_0^{\prime}(z_2) \nonumber \\
&+& \int\limits_{-\infty}^{\infty} \!\!\! dz \left[ \frac{\mu_1}{2} z \, \rho_1^{\prime}(z)
+ \mu_1 \, z^2 \, \rho_0^{\prime}(z) + \mu_{s,2} \, z \, \rho_0^{\prime}(z) \right] \,, \nonumber
\end{eqnarray} 
where $\mu_{s,2} \!=\! - \sigma \, \Delta \rho_1 / (\Delta \rho)^2
- 2 \delta \sigma / \Delta \rho$ \cite{Blokhuis93, Blokhuis06} with
$\Delta \rho_1 \!\equiv\! \rho_{1,\ell} - \rho_{1,v}$.
\vskip 10pt
\noindent
{\bf Cylindrical interface}
\vskip 5pt
\noindent
The analysis for the cylindrical interface is analogous
to that of the spherical interface. Following the same procedure as
for the spherical interface, the expressions for $\sigma$ and
$\delta \sigma$ in Eqs.(\ref{eq:sigma}) and (\ref{eq:delta})
are recovered and one obtains as an expression for the bending
rigidity $k$:
\begin{eqnarray}
\label{eq:rigidity}
k &=& \frac{1}{8} \int\limits_{-\infty}^{\infty} \!\!\! dz_1 \! \int \!\! d\vec{r}_{12} \;
U_{\rm att}(r) \, r^2 (1-s^2) \, \rho_0^{\prime}(z_1) \rho_1(z_2) \\
&+& \frac{1}{64} \int\limits_{-\infty}^{\infty} \!\!\! dz_1 \! \int \!\! d\vec{r}_{12} \;
U_{\rm att}(r) \, r^4 (1-s^2)^2 \, \rho_0^{\prime}(z_1) \rho_0^{\prime}(z_2) \nonumber \\
&+ & \int\limits_{-\infty}^{\infty} \!\!\! dz \left[ \frac{\mu_1}{4} z \, \rho_1^{\prime}(z)
+ \frac{\mu_1}{2} \, z^2 \, \rho_0^{\prime}(z) + 2 \mu_{c,2} \, z \, \rho_0^{\prime}(z) \right] \,, \nonumber
\end{eqnarray} 
where $\mu_{c,2} \!=\! - \sigma \, \Delta \rho_1 / (2 \, \Delta \rho)^2$
\cite{Blokhuis93, Blokhuis06}. An expression for the rigidity constant
associated with Gaussian curvature is then obtained by combining 
Eqs.(\ref{eq:k_sph}) and (\ref{eq:rigidity}):
\begin{eqnarray}
\label{eq:k_bar}
\bar{k} &=& - \frac{1}{4} \int\limits_{-\infty}^{\infty} \!\!\! dz_1 \! \int \!\! d\vec{r}_{12} \;
U_{\rm att}(r) \, r^2 (1-s^2) \, z_1^2 \, \rho_0^{\prime}(z_1) \rho_0^{\prime}(z_2) \\
&& - \frac{1}{96} \int\limits_{-\infty}^{\infty} \!\!\! dz_1 \! \int \!\! d\vec{r}_{12} \;
U_{\rm att}(r) \, r^4 (1-s^2) (1-5s^2) \, \rho_0^{\prime}(z_1) \rho_0^{\prime}(z_2) \nonumber \\
&& + (\mu_{s,2} - 4 \mu_{c,2}) \int\limits_{-\infty}^{\infty} \!\!\! dz \; z \, \rho_0^{\prime}(z) \,. \nonumber
\end{eqnarray} 
The expressions for $k$ and $\bar{k}$ differ in two ways somewhat from previous
expressions derived by us in ref.~\cite{Giessen98}. First, they are rewritten
in a more compact form with a printing error in ref.~\cite{Giessen98} corrected
(as noted by Barrett \cite{Barrett09}). Second, these expressions are derived without
reference to a particular choice for the location of the dividing surface, i.e.
for the location of the $z \!=\! 0$ plane. This feature allows us to investigate
the influence of the {\em choice} for the location of the dividing surface.
As already known, the surface tension and Tolman length are {\em independent}
of this choice but $k$ and $\bar{k}$ {\em do} depend on it. 
\vskip 10pt
\noindent
{\bf Choice for the location of the dividing surface}
\vskip 5pt
\noindent
We first consider the density profile of the {\em planar} interface, obtained by
solving the differential equation in Eq.(\ref{eq:EL_planar}), to investigate the
consequences of the choice for the location of the dividing surface for $\delta$
and $\bar{k}$. One may verify that when $\rho_0(z)$ is a particular solution of
the differential equation in Eq.(\ref{eq:EL_planar}), then the {\em shifted}
density profile
\begin{equation}
\rho_0(z) \longrightarrow \rho_0(z-z_0) \,,
\end{equation}
is also a solution for arbitrary value of the integration constant $z_0$. However,
since the expressions for $\delta$ and $\bar{k}$ feature $z$ (or $z_1$) in
the integrand, such a shift has consequences for the different contributions
to $\delta$ and $\bar{k}$. To investigate this in more detail, we first
place the dividing surface of the planar system at the equimolar surface,
$z \!=\! z_e$, which is defined such that the excess density is zero \cite{Gibbs}:
\begin{equation}
\label{eq:equimolar}
\int\limits_{-\infty}^{\infty} \!\!\! dz \; [ \rho_0(z) - \rho_{0,\ell} \, \Theta(z_e-z) - \rho_{0,v} \, \Theta(z-z_e) ]
= - \int\limits_{-\infty}^{\infty} \!\!\! dz \; (z-z_e) \, \rho_0^{\prime}(z) = 0 \,,
\end{equation}
where $\Theta(z)$ is the Heaviside function. When all distances to the surface
are measured with respect to the equimolar plane, we need to replace $z$
by $z-z_e$ in the expressions for $\delta$ and $\bar{k}$. For the
Tolman length in Eq.(\ref{eq:delta}) we then find that:
\begin{equation}
\label{eq:delta_equimolar}
\delta \sigma = \frac{1}{4} \int\limits_{-\infty}^{\infty} \!\!\! dz_1 \! \int \!\! d\vec{r}_{12} \;
U_{\rm att}(r) \, r^2 (1-s^2) \, (z_1 - z_e) \, \rho_0^{\prime}(z_1) \rho_0^{\prime}(z_2) \,,
\end{equation}
where we have used Eq.(\ref{eq:equimolar}). Now, to investigate the consequences of
shifting the dividing surface away from the equimolar surface by a distance $\Delta$,
we replace $z \!\rightarrow\! z - (z_e + \Delta)$ in the expression for
the Tolman length in Eq.(\ref{eq:delta}). One may easily verify that on account of
the fact that $\mu_1 \!=\! 2 \sigma / \Delta \rho$ the Tolman length then again reduces
to the expression in Eq.(\ref{eq:delta_equimolar}) which proofs that the Tolman
length is {\em independent} of the choice for the location of the dividing surface.

Replacing $z \!\rightarrow\! z - z_e$ in the expression for the rigidity
constant associated with Gaussian curvature in Eq.(\ref{eq:k_bar}), we find
that $\bar{k}$ simplifies to
\begin{eqnarray}
\label{eq:k_bar_equimolar}
\bar{k}_{\rm equimolar} &=& - \frac{1}{4} \int\limits_{-\infty}^{\infty} \!\!\! dz_1 \! \int \!\! d\vec{r}_{12} \;
U_{\rm att}(r) \, r^2 (1-s^2) \, (z_1-z_e)^2 \, \rho_0^{\prime}(z_1) \rho_0^{\prime}(z_2) \\
&& - \frac{1}{96} \int\limits_{-\infty}^{\infty} \!\!\! dz_1 \! \int \!\! d\vec{r}_{12} \;
U_{\rm att}(r) \, r^4 (1-s^2) (1-5s^2) \, \rho_0^{\prime}(z_1) \rho_0^{\prime}(z_2) \,. \nonumber
\end{eqnarray}
Again, we may investigate the consequence of shifting the dividing surface
by replacing $z \!\rightarrow\! z - (z_e + \Delta)$ in the expression for
$\bar{k}$ in Eq.(\ref{eq:k_bar}). We then find that
\begin{equation}
\bar{k} = \bar{k}_{\rm equimolar} + \sigma \, \Delta^2 \,.
\end{equation}
This equation shows that $\bar{k}$ {\em does} depend on the choice for
the location of the dividing surface. It also shows that $\bar{k}$
evaluated for the {\em equimolar} surface ($\Delta \!=\! 0$), corresponds
to the {\em lowest} possible value for $\bar{k}$ and is the {\em least}
sensitive to a shift in the location of the dividing surface.

To address the influence of the dividing surface on the value of the
{\em bending rigidity} $k$, we need to consider the properties of the
density profile $\rho_1(z)$ as well. One may verify that when $\rho_1(z)$
is a particular solution of Eq.(\ref{eq:EL_1}) then also
\begin{equation}
\label{eq:rho_1}
\rho_1(z) \longrightarrow \rho_1(z) + \alpha \, \rho_0^{\prime}(z) \,,
\end{equation}
is a solution for arbitrary value of the integration constant $\alpha$.
Now, one may easily verify by inserting Eq.(\ref{eq:rho_1}) into
Eq.(\ref{eq:rigidity}) that $k$ is {\em independent} of the value
of the integration constant. This means that just like $\delta$ and $\bar{k}$
we only need to consider the influence of the choice for the location of
the dividing surface of the {\em planar} density profile $\rho_0(z)$.
For the {\em equimolar} surface, the expression for the bending rigidity
in Eq.(\ref{eq:rigidity}) reduces to:
\begin{eqnarray}
\label{eq:k_equimolar}
k_{\rm equimolar} &=& \frac{1}{8} \int\limits_{-\infty}^{\infty} \!\!\! dz_1 \! \int \!\! d\vec{r}_{12} \;
U_{\rm att}(r) \, r^2 (1-s^2) \, \rho_0^{\prime}(z_1) \rho_1(z_2) \\
&+& \frac{1}{64} \int\limits_{-\infty}^{\infty} \!\!\! dz_1 \! \int \!\! d\vec{r}_{12} \;
U_{\rm att}(r) \, r^4 (1-s^2)^2 \, \rho_0^{\prime}(z_1) \rho_0^{\prime}(z_2) \nonumber \\
&+& \frac{\mu_1}{4} \int\limits_{-\infty}^{\infty} \!\!\! dz \left[ (z - z_e) \, \rho_1^{\prime}(z)
+ 2 \, (z-z_e)^2 \, \rho_0^{\prime}(z) \right] \,. \nonumber
\end{eqnarray}
Shifting the dividing surface by replacing $z \!\rightarrow\! z - (z_e + \Delta)$
in the expression for $k$ in Eq.(\ref{eq:rigidity}), we then find that
\begin{equation}
k = k_{\rm equimolar} - \sigma \, \Delta^2 \,.
\end{equation}
It is concluded that also the bending rigidity $k$ {\em does} depend on the
choice for the location of the dividing surface. The bending rigidity
evaluated for the {\em equimolar} surface ($\Delta \!=\! 0$), now corresponds
to the {\em largest} possible value for $k$ but it is again the {\em least}
sensitive to a shift in the location of the dividing surface.

\begin{figure}
\centering
\includegraphics[angle=270,width=250pt]{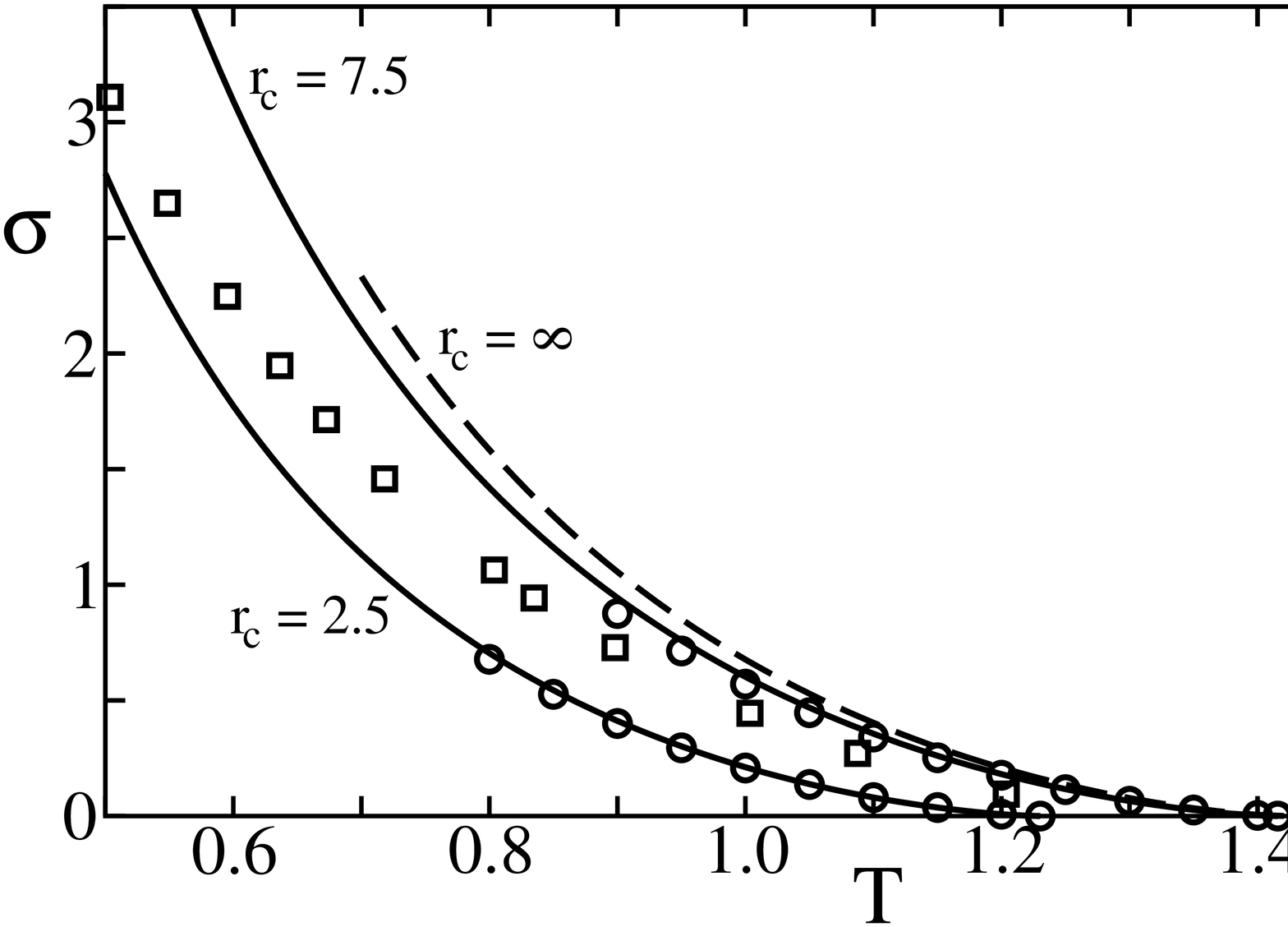}
\includegraphics[angle=270,width=250pt]{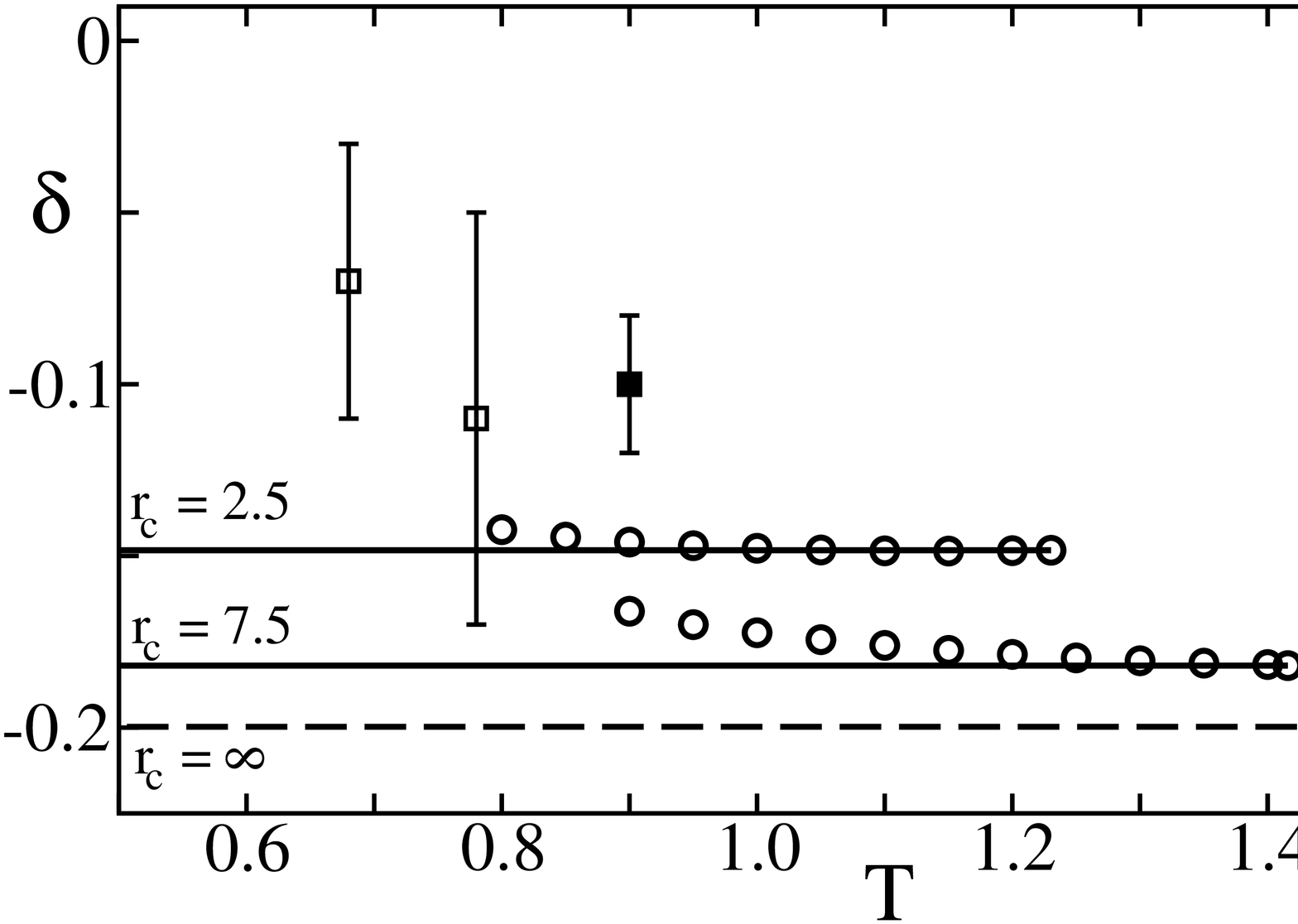}
\caption{Surface tension $\sigma$ (in units of $k_{\rm B} T / d^2$) and 
Tolman length $\delta$ (in units of $d$) as a function of reduced temperature.
Circular symbols are the results of the full DFT calculations in
Eqs.(\ref{eq:sigma}) and (\ref{eq:delta}). The solid lines are the
squared-gradient approximations of Section \ref{sec-explicit}.
Square symbols are simulation results for $\sigma$ from ref.~\cite{Baidakov07}
and for $\delta$ from ref.~\cite{Giessen09} (solid square) and
ref.~\cite{Binder10} (two open squares).}
\label{Fig:sigma_delta}
\end{figure}

\vskip 10pt
\noindent
The procedure to determine the curvature coefficients $\sigma$, $\delta$,
$k$ and $\bar{k}$ is now as follows. The planar profile $\rho_0(z)$ is first
determined from the differential equation in Eq.(\ref{eq:EL_planar}) with
$\rho_{0,\ell}$, $\rho_{0,v}$, $\mu_{\rm coex}$ and $p_{\rm coex}$
derived from solving the set of equations in Eq.(\ref{eq:bulk_0}). From
$\rho_0(z)$, the location of the equimolar plane $z \!=\! z_e$ is determined
from Eq.(\ref{eq:equimolar}) and the curvature coefficients $\sigma$, $\delta$
and $\bar{k}$ are evaluated from the integrals in Eq.(\ref{eq:sigma}),
(\ref{eq:delta_equimolar}) and (\ref{eq:k_bar_equimolar}), respectively.
The constant $\mu_1$ is subsequently determined from $\mu_1 \!=\! 2 \sigma
/ \Delta \rho$ which allows us to determine the bulk density values
$\rho_{1,\ell/v}$ from $\rho_{1,\ell/v} \!=\! \mu_1 / f^{\prime \prime}
(\rho_{0,\ell/v})$. For given $\rho_0(z)$ and $\mu_1$, the differential
equation for $\rho_1(z)$ in Eq.(\ref{eq:EL_1}) is solved with the
boundary conditions $\rho_1(-\infty) \!=\! \rho_{1,\ell}$ and
$\rho_1(\infty) \!=\! \rho_{1,v}$. Finally, with $\rho_1(z)$ determined,
$k$ can be evaluated from the integral in Eq.(\ref{eq:k_equimolar}).

\begin{figure}
\centering
\includegraphics[angle=270,width=250pt]{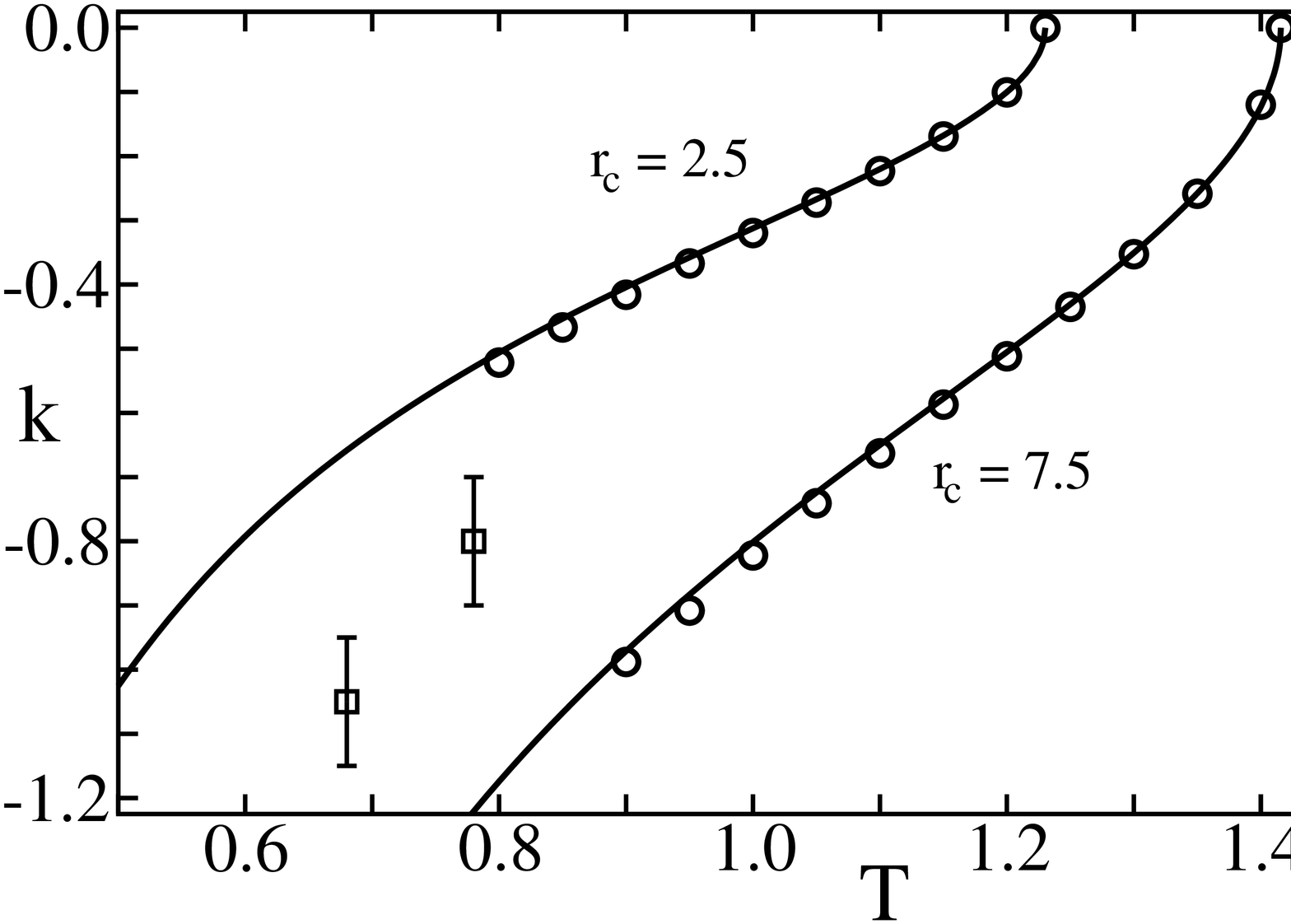}
\includegraphics[angle=270,width=250pt]{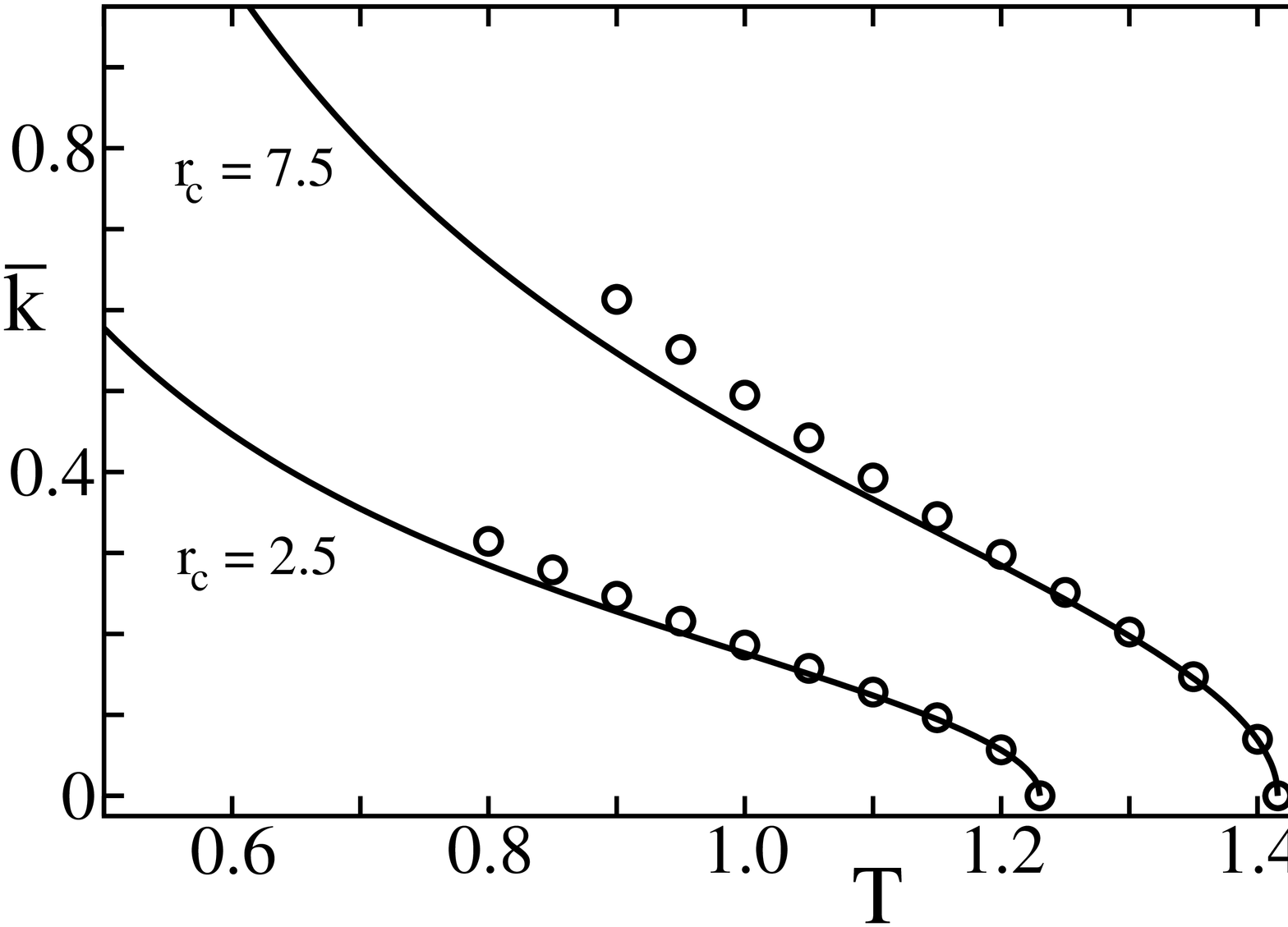}
\includegraphics[angle=270,width=250pt]{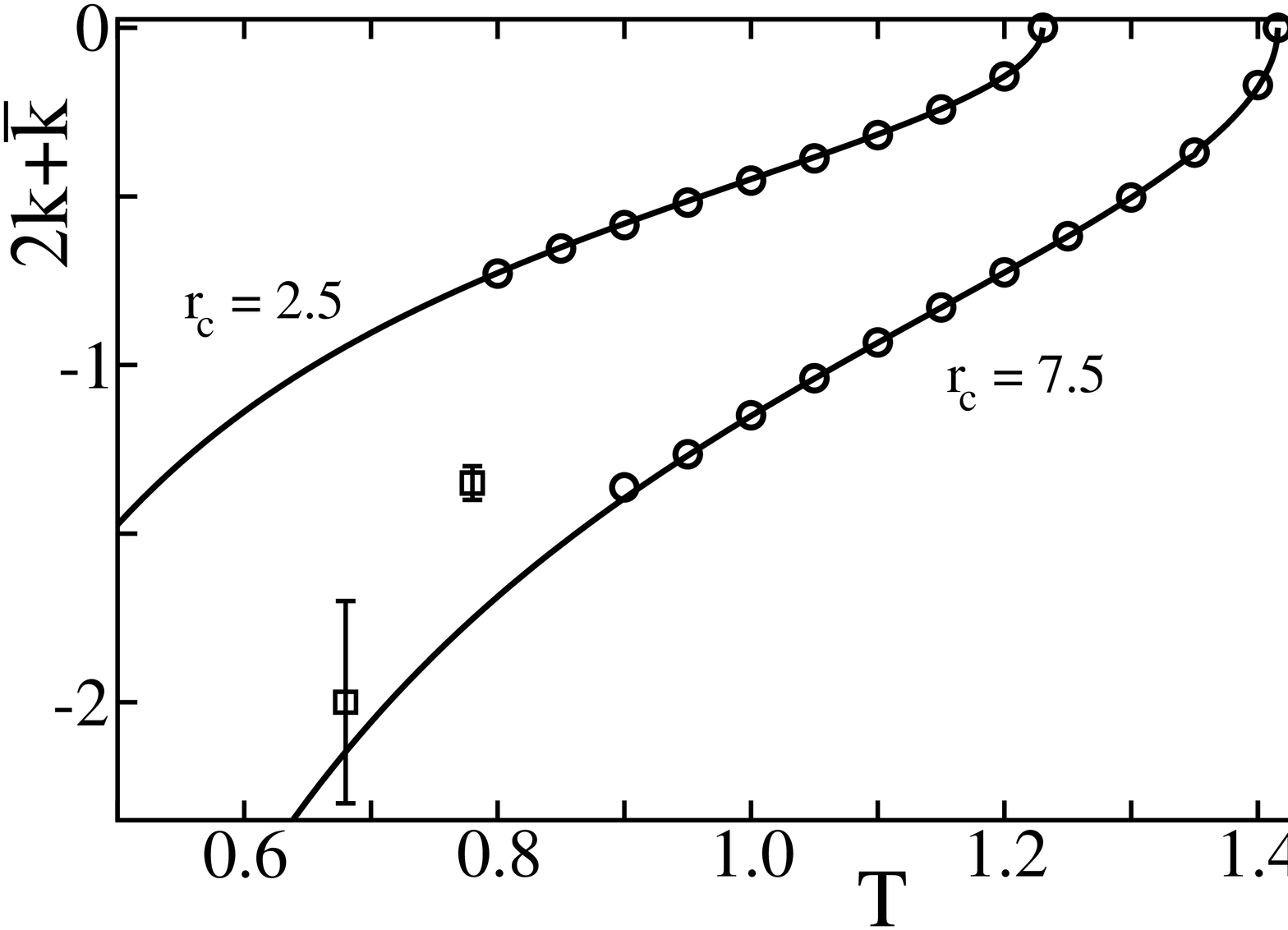}
\caption{Bending rigidity $k$, Gaussian rigidity $\bar{k}$,
and the combination $2 k + \bar{k}$ (in units of $k_{\rm B} T$)
as a function of temperature. The rigidity constants are evaluated
using the {\em equimolar} surface as the dividing surface. Circular symbols
are the results of the full DFT calculations in Eqs.(\ref{eq:k_bar_equimolar})
and (\ref{eq:k_equimolar}). The solid lines are the squared-gradient
approximations of Section \ref{sec-explicit}. Square symbols are simulation
results from ref.~\cite{Binder10}.}
\label{Fig:kabar_ka}
\end{figure}

This procedure is carried out (again) using the cut-off and shifted
Lennard-Jones potential in Eq.(\ref{eq:LJ}) for the attractive part
of the interaction potential. Figure \ref{Fig:sigma_delta} shows the
surface tension and Tolman length as a function of temperature.
The circular symbols are the values for $\sigma$ and $\delta$ calculated
using DFT for two values of the LJ cut-off radius $r_c$. The solid lines
are the squared-gradient approximations in Section \ref{sec-explicit}
for $r_c \!=$ 2.5, 7.5, and $\infty$. As square symbols, we show
computer simulation results for $\sigma$ from ref.~\cite{Baidakov07},
the single simulation result for $\delta$ from ref.~\cite{Giessen09}
(solid square) and results for $\delta$ from simulations by the group
of Binder \cite{Binder10} (open squares).

In Figure \ref{Fig:kabar_ka}, the bending rigidity $k$, Gaussian rigidity
$\bar{k}$, and the combination $2 k + \bar{k}$ are shown as a function of
temperature. The rigidity constants are evaluated using the {\em equimolar}
surface for the location of the dividing surface. The circular symbols are
the values for $k$ and $\bar{k}$ calculated using DFT for two values of the
reduced LJ cut-off radius $r_c \!=$ 2.5 and 7.5, with the solid lines the 
corresponding squared-gradient approximations determined in the next Section.
Also shown are simulations results by the group of Binder \cite{Binder10}.
Although a detailed comparison of the DFT and simulation results is not
really appropriate due to a difference in cut-off used, the agreement in
sign and order of magnitude is rather satisfactory.

\section{Squared-gradient expressions}
\label{sec-explicit}

\noindent
The evaluation of $\delta$, $k$ and $\bar{k}$ requires the full numerical
evaluation of the density profiles $\rho_0(z)$ and $\rho_1(z)$ from
the differential equations in Eqs.(\ref{eq:EL_planar}) and (\ref{eq:EL_1}).
This procedure is quite elaborate, prompting a need for simple formulas
that provide (approximate) numbers for the various coefficients.
In this section we provide a rather accurate approximation scheme based
on the squared-gradient approximation which only requires the calculation
of the phase diagram as input.

The squared-gradient theory for surfaces dates back to the work of van der
Waals in 1893 \cite{vdW}. Its free energy functional is derived from
Eq.(\ref{eq:Omega_DFT}) by assuming that gradients in the density are
small so that $\rho(\vec{r}_2)$ may be expanded around $\rho(\vec{r}_1)$.
This leads to:
\begin{equation}
\label{eq:Omega_SQ}
\Omega[\rho] = \int \!\! d\vec{r} \; \left[ m \, | \vec{\nabla} \rho(\vec{r}) |^2
+ f(\rho) - \mu \rho(\vec{r}) \right] \,,
\end{equation} 
where the squared-gradient coefficient $m$ is given by
\begin{equation}
\label{eq:m}
m \equiv - \frac{1}{12} \int \!\! d\vec{r}_{12} \; r^2 \, U_{\rm att}(r) \,.
\end{equation}
Expressions for the curvature coefficients in squared-gradient theory were
formulated some time ago. For the surface tension of the planar
interface, we have the familiar expression given by van der Waals \cite{vdW}:
\begin{equation}
\label{eq:sigma_SQ}
\sigma = 2 \, m  \int\limits_{-\infty}^{\infty} \!\!\! dz \; \rho_0^{\prime}(z)^2 \,.
\end{equation}
For the Tolman length, Fisher and Wortis derived the following expression \cite{FW}:
\begin{equation}
\label{eq:delta_SQ}
\delta \sigma = - 2 \, m \int\limits_{-\infty}^{\infty} \!\!\! dz \; (z - z_e) \, \rho_0^{\prime}(z)^2 \,.
\end{equation}
For the bending and Gaussian rigidity, one has \cite{Blokhuis93}:
\begin{eqnarray}
\label{eq:k_SQ}
k &=& - m \int\limits_{-\infty}^{\infty} \!\!\! dz \; \rho_0(z) \, \rho_1^{\prime}(z)
+ \int\limits_{-\infty}^{\infty} \!\!\! dz \left[ \frac{\mu_1}{4} z \, \rho_1^{\prime}(z)
+ \frac{\mu_1}{2} \, z^2 \, \rho_0^{\prime}(z) + 2 \mu_{c,2} \, z \, \rho_0^{\prime}(z) \right] \,, \nonumber \\
\bar{k} &=& 2 \, m \int\limits_{-\infty}^{\infty} \!\!\! dz \; z^2 \, \rho_0^{\prime}(z)^2
+ (\mu_{s,2} - 4 \mu_{c,2}) \int\limits_{-\infty}^{\infty} \!\!\! dz \; z \, \rho_0^{\prime}(z) \,,
\end{eqnarray}
which, evaluated using the equimolar surface for the location of the dividing
surface, reduce to:
\begin{eqnarray}
\label{eq:k_equimolar_SQ}
k_{\rm equimolar} &=& - m \int\limits_{-\infty}^{\infty} \!\!\! dz \; \rho_0(z) \, \rho_1^{\prime}(z)
+ \frac{\mu_1}{4} \int\limits_{-\infty}^{\infty} \!\!\! dz \left[ (z - z_e) \, \rho_1^{\prime}(z)
+ 2 \, (z-z_e)^2 \, \rho_0^{\prime}(z) \right] \,, \nonumber \\
\bar{k}_{\rm equimolar} &=& 2 \, m \int\limits_{-\infty}^{\infty} \!\!\! dz \; (z-z_e)^2 \, \rho_0^{\prime}(z)^2 \,.
\end{eqnarray}
To evaluate these expressions, the density profiles $\rho_0(z)$ and $\rho_1(z)$
still need to be determined from the expanded Euler-Lagrange equation:
\begin{eqnarray}
\label{eq:EL_SQ_0}
f^{\prime}(\rho_0) &=& \mu_{\rm coex} + 2 m \, \rho_0^{\prime\prime}(z) \,, \\
\label{eq:EL_SQ_1}
f^{\prime\prime}(\rho_0) \, \rho_1(z) &=& \mu_1 + 2 m \, \rho^{\prime\prime}_1(z) + 4 m \, \rho^{\prime}_0(z) \,.
\end{eqnarray}
In order to solve these equations, it is useful to assume proximity to the critical
point so that the free energy density may be approximated by the usual double-well form:
\begin{equation}
\label{eq:rho^4}
f(\rho) - \mu_{\rm coex} \rho + p_{\rm coex} = \frac{m}{(\Delta \rho)^2 \, \xi^2} \,
(\rho - \rho_{0,\ell})^2 \, (\rho - \rho_{0,v})^2 \,,
\end{equation}
where the bulk correlation length $\xi$ is related to the second derivative
of $f(\rho)$ evaluated at either bulk density. Solving the Euler-Lagrange
equation in Eq.(\ref{eq:EL_SQ_0}) then leads to the usual $tanh$-form for
the planar density profile \cite{RW}:
\begin{equation}
\label{eq:tanh}
\rho_0(z) = \frac{1}{2} ( \rho_{0,\ell} + \rho_{0,v} ) - \frac{\Delta \rho}{2} \, \tanh((z-z_e)/2\xi) \,.
\end{equation}
One may verify that solving the Euler-Lagrange equation in Eq.(\ref{eq:EL_SQ_1})
gives the following general solution for $\rho_1(z)$ \cite{Blokhuis93}:
\begin{equation}
\label{eq:rho_1_SQ}
\rho_1(z) = \frac{1}{3} \, m \, (\Delta \rho)^2 \, \xi + \alpha \, \rho_0^{\prime}(z) \,.
\end{equation}
As already discussed, the rigidity constant is {\em independent} of the integration
constant $\alpha$. Inserting these profiles into the expressions for $\sigma$, $k$
and $\bar{k}$ in Eqs.(\ref{eq:sigma_SQ}) and (\ref{eq:k_equimolar_SQ}), one finds
\cite{Blokhuis93}:
\begin{eqnarray}
\label{eq:k_kbar_SQ}
\sigma &=& \frac{m \, (\Delta \rho)^2}{3 \, \xi} \,, \\
k_{\rm equimolar} &=& - \frac{1}{9} (\pi^2 - 3) \, m \, (\Delta \rho)^2 \, \xi \,, \nonumber \\
\bar{k}_{\rm equimolar} &=& \frac{1}{9} (\pi^2 - 6) \, m \, (\Delta \rho)^2 \, \xi \,. \nonumber
\end{eqnarray}
For the symmetric double-well form for $f(\rho)$, the Tolman length
is identically zero. To obtain an estimate for $\delta$ it is
therefore necessary to consider leading order corrections to the
double-well form for $f(\rho)$ in Eq.(\ref{eq:rho^4}) \cite{FW, Giessen98}.
This leads to the following (constant) value for the Tolman length \cite{Giessen98}:
\begin{equation}
\label{eq:delta_SQ_cp}
\delta = -0.286565 \, \sqrt{m / a} \,.
\end{equation}
The prefactor depends on the precise form for $f(\rho)$ and the number
quoted is specific to the Carnahan-Starling equation of state \cite{delta_c}.

All these formulas are derived assuming proximity to the critical point,
but it turns out that they also provide a good approximation in a wide
temperature range when the value of $\xi$ is chosen judiciously.
This is done by using the fact that in squared-gradient theory the
surface tension $\sigma$ may be determined from $f(\rho)$ {\em directly}
without the necessity to determine the density profile $\rho_0(z)$ \cite{RW}:
\begin{equation}
\label{eq:sigma_SQ_cp}
\sigma = 2 \, \sqrt{m}  \int\limits_{\rho_{0,v}}^{\rho_{0,\ell}} \!\! d\rho \;
\sqrt{f(\rho) - \mu_{\rm coex} \, \rho + p_{\rm coex}} \,.
\end{equation}
An effective value for $\xi$ may now be chosen such that the two expressions
for the surface tension in Eqs.(\ref{eq:k_kbar_SQ}) and (\ref{eq:sigma_SQ_cp})
are equal. This gives for $\xi$:
\begin{equation}
\label{eq:xi}
\xi \longrightarrow \xi_{\rm eff} \equiv \frac{m \, (\Delta \rho)^2}{3 \, \sigma} \,.
\end{equation}
with $\sigma$ given by Eq.(\ref{eq:sigma_SQ_cp}).

The procedure to determine the solid lines in Figures \ref{Fig:sigma_delta}
and \ref{Fig:kabar_ka} is now as follows. For a certain interaction potential,
such as the Lennard-Jones potential in Eq.(\ref{eq:LJ}), the interaction
parameters $a$ and $m$ are calculated. Next, as a function of temperature,
the bulk thermodynamic variables $\rho_{0,\ell}$, $\rho_{0,v}$, $\mu_{\rm coex}$
and $p_{\rm coex}$ are derived from solving the set of equations in Eq.(\ref{eq:bulk_0}).
The surface tension is then calculated from Eq.(\ref{eq:sigma_SQ_cp})
and $\xi$ from Eq.(\ref{eq:xi}). With all parameters known, the
curvature coefficients are finally calculated from Eqs.(\ref{eq:k_kbar_SQ})
and (\ref{eq:delta_SQ_cp}).

\section{Long-ranged interactions: dispersion forces}
\label{sec-dispersion}

\noindent
The surface tension, Tolman length and rigidity constants have all been explicitly
evaluated using a Lennard-Jones potential that is cut-off beyond a certain distance $r_c$.
In this section we address the consequences of using the {\em full} Lennard-Jones
potential. It is easily verified that the phase diagram in Figure \ref{Fig:pd}
remains essentially the same when the cut-off is changed from $r_c \!=\! 7.5$ to
$r_c \!=\! \infty$, but that the shift in surface tension and Tolman length
is increasingly noticeable (see Figure \ref{Fig:sigma_delta}).
An inspection of the explicit expressions for the rigidity constants in
Eqs.(\ref{eq:rigidity}) and (\ref{eq:k_bar}) teaches us that both $k$ and $\bar{k}$
{\em diverge} when $r_c$ increases to infinity \cite{Blokhuis92a, Dietrich}.
This divergence is an indication that the expansion of the free energy is no
longer of the form in Eq.(\ref{eq:sigma_s(R)}) or (\ref{eq:sigma_c(R)}), and
it has to be replaced by 
\begin{eqnarray}
\label{eq:sigma_s_LR}
\sigma_s(R) &=& \sigma - \frac{2 \delta \sigma}{R} + (2 k_s + \bar{k}_s) \, \frac{\log(d/R)}{R^2} + \ldots \\
\label{eq:sigma_c_LR}
\sigma_c(R) &=& \sigma - \frac{\delta \sigma}{R} + k_s \,\frac{\log(d/R)}{2 R^2} + \ldots
\end{eqnarray}
where the dots represent terms of ${\cal O}(1/R^2)$. The coefficients of the logarithmic
terms may be extracted from the expressions for $k$ and $\bar{k}$ in Eqs.(\ref{eq:rigidity})
and (\ref{eq:k_bar}). They depend on the tail of the interaction potential, but are
otherwise quite universal:
\begin{eqnarray}
\label{eq:k_s}
k_s &=& \frac{\pi}{8} \, \varepsilon \, d^6 \, (\Delta \rho)^2 \,, \\
\label{eq:k_bar_s}
\bar{k}_s &=& - \frac{\pi}{12} \, \varepsilon \, d^6 \, (\Delta \rho)^2 \,.
\end{eqnarray}
This expression for $k_s$ is equal to that obtained in a DFT analysis of the
singular part of the wave vector dependent surface tension of the fluctuating
interface \cite{Blokhuis09}. These expressions can also be derived from virial
expressions for the rigidity constants when a sharp-kink approximation
\cite{Dietrich} is made for the density profile \cite{correctie_Blokhuis92a}.
The form for $2k_s + \bar{k}_s$ obtained by combining Eqs.(\ref{eq:k_s})
and (\ref{eq:k_bar_s}) was first derived by Hooper and Nordholm in
ref.~\cite{Hooper}.

\begin{figure}
\centering
\includegraphics[angle=270,width=250pt]{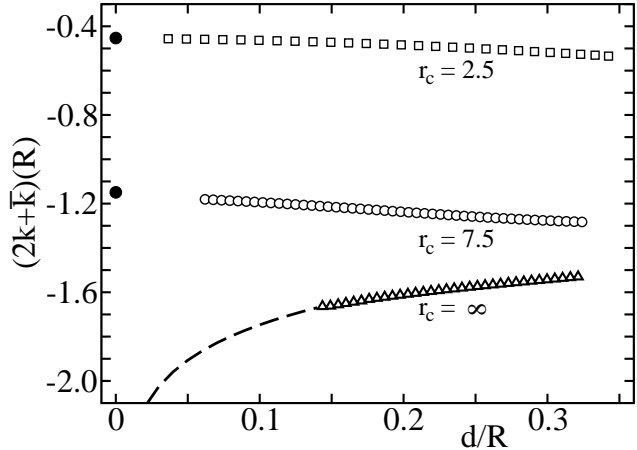}
\caption{The combination $(2 k + \bar{k})(R)$ (in units of $k_{\rm B} T$) as defined
by Eq.(\ref{eq:2k_k_bar_LR}) as a function of the reciprocal equimolar radius $d/R$.
The symbols are the results of DFT calculations at reduced temperature $T^* \!=\!$ 1.0
and three values of the reduced LJ cut-off $r_c \!=$ 2.5, 7.5 and $\infty$.
Solid circles are the corresponding values for $2 k + \bar{k}$ calculated from
Eqs.(\ref{eq:k_bar_equimolar}) and (\ref{eq:k_equimolar}). The dashed line
is the curve $\pi/(6 T^*) (\Delta \rho^*)^2 \log(R_0/R)$ with $R_0 \!\simeq$ 0.005 $d$.}
\label{Fig:k_sph_R}
\end{figure}

To demonstrate the divergence of the second order term in Eq.(\ref{eq:sigma_s_LR}),
the surface tension of a spherical liquid droplet as a function of the radius
is determined for three values of the reduced LJ cut-off radius $r_c \!=$ 2.5,
7.5 and $r_c \!=\! \infty$. The regular contributions to $\sigma_s(R)$ from
$\sigma$ and $\delta$ are subtracted, so that we may define
\begin{equation}
\label{eq:2k_k_bar_LR}
(2 k + \bar{k})(R) \equiv \left (\sigma_s(R) - \sigma \right) \, R^2 + 2 \delta \sigma \, R \,.
\end{equation}
This quantity is defined such that when the expansion in Eq.(\ref{eq:sigma_s(R)})
for short-ranged forces is inserted, it reduces to $2 k + \bar{k}$ in the limit
that $R \rightarrow \infty$. For long-ranged forces ($r_c \!=\! \infty$), insertion
of Eq.(\ref{eq:sigma_s_LR}) into Eq.(\ref{eq:2k_k_bar_LR}) gives a logarithmic
divergence in this limit. This is verified by the DFT calculations shown in
Figure \ref{Fig:k_sph_R} as the various symbols. For $r_c \!=$ 2.5 and $r_c \!=$ 7.5,
the results indeed tend to the values obtained from the direct evaluation
of $2 k + \bar{k}$ using Eqs.(\ref{eq:k_bar_equimolar}) and (\ref{eq:k_equimolar})
(solid circles). For $r_c \!=\! \infty$ (triangular symbols) a slight
divergence can be made out. This divergence is consistent with the dashed
line, which is the divergence as described by combining the coefficients
in Eqs.(\ref{eq:k_s}) and (\ref{eq:k_bar_s}).

\section{Discussion}
\label{sec-discussion}

\noindent
In the context of density functional theory, we have shown that the surface
tension of a spherical liquid droplet as a function of its inverse radius
is well-represented by a parabola with its second derivative related to the
rigidity constants $k$ and $\bar{k}$. Compact formulas for the evaluation
of $k$ and $\bar{k}$ are derived in terms of the density profiles $\rho_0(z)$
and $\rho_1(z)$, which are in line with previous formulas presented by us
\cite{Giessen98} and by Barrett \cite{Barrett09}. A number of conclusions
can be made with regard to these formulas:

\begin{itemize}

\item{The rigidity constants $k$ and $\bar{k}$ depend on the choice for the
location of the dividing surface of the planar density profile $\rho_0(z)$.
This dependency reflects the fact that when the location of the radius $R$
is chosen differently, the curve of $\sigma_s(R)$ versus $1/R$ changes
somewhat and the second derivative ($2 k + \bar{k}$) naturally needs to be
amended.}

\item{The most natural choice for a one-component system, is to locate
the dividing surface of the planar interface according to the {\em equimolar
surface}. For this choice both $k$ and $\bar{k}$ are the {\em least} sensitive
to a change in the location of the dividing surface. Furthermore, the equimolar
value for $k$ corresponds to its {\em maximum} value and the equimolar
value for $\bar{k}$ corresponds to its {\em minimum} value.}

\item{The bending rigidity $k$ depends on the density profile $\rho_1(z)$,
which measures the extent by which molecules rearrange themselves when the
interface is curved. The bending rigidity is, however, {\em independent}
of the choice made for the location of the dividing surface of $\rho_1(z)$
(value of $\alpha$ in Eq.(\ref{eq:rho_1})) \cite{constraints}.}

\end{itemize}

\noindent
Using a cut-off and shifted Lennard-Jones potential for the attractive
part of the interaction potential, the Tolman length and rigidity constants
have been calculated with the result that $\delta$ is negative with
a value of minus 0.1-0.2 $d$, $k$ is also negative with a value around
minus 0.5-1.0 $k_{\rm B} T$, and $\bar{k}$ is positive with a value of
a bit more than half the magnitude of $k$. It is not expected that
these results depend sensitively on the type of density functional
theory used and we have shown that even an approximation scheme based
on squared-gradient theory is {\em quantitatively} accurate.

Our DFT results are expected to give an accurate {\em qualitative}
description of the rigidity constants determined in experiments
or computer simulations. First results of computer simulations by
the group of Binder \cite{Binder10} shown in Figure \ref{Fig:kabar_ka},
seem to support this expectation, but further computer simulations
are necessary. The agreement should cease to exist close to the critical
point, however. Since the DFT calculations are all mean-field in character,
the critical exponents obtained for both rigidity constants are the
mean-field values of $1/2$, which indicates that both $k$ and $\bar{k}$
are zero at $T_c$. Although it has not been proved rigorously, one
expects that in reality the rigidity constants are finite at the
critical point $k$, $\bar{k} \!\propto\! k_{\rm B} T_c$. The situation
is somewhat more subtle for the rigidity constant associated with the
description of {\em surface fluctuations}. Then, the bending rigidity
is again negative but it vanishes on approach to the critical point
with the same exponent as the surface tension \cite{Blokhuis09}.

The inspection of the explicit expressions presented for the
rigidity constants is the most convincing method to investigate
the possible presence of logarithmic corrections \cite{Henderson92,
Rowlinson94, Fisher}, to replace the rigidity constants. For
short-ranged interactions between molecules, the rigidity constants
are definitely {\em finite}, but for an interaction potential
that falls of as $1/r^6$ for large intermolecular distances
(dispersion forces), the rigidity constants are {\em infinite}
indicating that the $1/R^2$ term in the expansion of the surface
tension needs to be replaced by a logarithmic term proportional
to $\log(R) / R^2$. The proportionality constants of the logarithmic
corrections are found to be quite universal since they probe the
systems long-distance behaviour and are in agreement with previous
analyses \cite{Blokhuis92a, Hooper, Dietrich, correctie_Blokhuis92a}.

\vskip 15pt
\noindent
{\Large\bf Acknowledgment}
\vskip 5pt
\noindent
A.E.v.G. acknowledges the generous support from an American Chemical
Society Petroleum Research Fund.


\appendix
\section{Alternative DFT expressions}
\setcounter{equation}{0}

\noindent
It may be useful to re-express the curvature coefficients $\delta$, $k$,
and $\bar{k}$ such that any reference to the chemical potential is absent.
For the Tolman length the expression for $\mu_1$ in Eq.(\ref{eq:EL_1})
may be used to rewrite Eq.(\ref{eq:delta}) as:
\begin{equation}
\label{eq:delta_alternative}
\delta \sigma = \frac{1}{4} \int\limits_{-\infty}^{\infty} \!\!\! dz_1 \! \int \!\! d\vec{r}_{12} \;
U_{\rm att}(r) \, r^2 (1-s^2) \, \rho_0^{\prime}(z_1) \rho_1^{\prime}(z_2) \,.
\end{equation}
This expression is quite useful since it can be used to verify that the
density profile $\rho_1(z)$ determined numerically by solving the differential
equation in Eq.(\ref{eq:EL_1}), leads to the same value for the Tolman length
when evaluated using Eq.(\ref{eq:delta}).

In order to transform the rigidity constants in a similar manner, we first need
to expand the Euler-Lagrange equation in Eq.(\ref{eq:EL_sphere}) to {\em second
order} in $1/R$. For the spherical interface, one finds:
\begin{eqnarray}
\label{eq:EL_2_s}
\mu_{s,2} &=& f^{\prime\prime}_{\rm hs}(\rho_0) \, \rho_{s,2}(z_1) 
+ \frac{1}{2} f^{\prime\prime\prime}_{\rm hs}(\rho_0) \, \rho_1(z_1)^2
+ \int \!\! d\vec{r}_{12} \; U_{\rm att}(r) \, [ \, \rho_{s,2}(z_2) \\
&+& \frac{r^2}{2} (1-s^2) \, \rho^{\prime}_1(z_2) 
- \frac{r^2}{2} (1-s^2) \, z_2 \, \rho^{\prime}_0(z_2) 
+ \frac{r^4}{8} (1-s^2)^2 \, \rho^{\prime\prime}_0(z_2) \, ] \,. \nonumber
\end{eqnarray}
The analogous expansion for the cylindrical interface gives:
\begin{eqnarray}
\label{eq:EL_2_c}
\mu_{c,2} &=& f^{\prime\prime}_{\rm hs}(\rho_0) \, \rho_{c,2}(z_1)  
+ \frac{1}{8} f^{\prime\prime\prime}_{\rm hs}(\rho_0) \, \rho_1(z_1)^2
+ \int \!\! d\vec{r}_{12} \; U_{\rm att}(r) \, [ \, \rho_{c,2}(z_2) \\
&+& \frac{r^2}{8} (1-s^2) \, \rho^{\prime}_1(z_2) 
- \frac{r^2}{4} (1-s^2) \, z_2 \, \rho^{\prime}_0(z_2) 
+ \frac{3 r^4}{64} (1-s^2)^2 \, \rho^{\prime\prime}_0(z_2) \, ] \,. \nonumber
\end{eqnarray}
Inserting these expressions for $\mu_{s,2}$ and $\mu_{c,2}$ into
Eqs.(\ref{eq:rigidity}) and (\ref{eq:k_bar}), one finds:
\begin{eqnarray}
\label{eq:k_&_k_bar_alternative}
k &=& - \int\limits_{-\infty}^{\infty} \!\!\! dz_1 \! \int \!\! d\vec{r}_{12} \;
U_{\rm att}(r) \, r^2 (1-s^2) \, \rho_0^{\prime}(z_1) \rho_{c,2}^{\prime}(z_2) \\
&& - \frac{1}{8} \int\limits_{-\infty}^{\infty} \!\!\! dz_1 \! \int \!\! d\vec{r}_{12} \;
U_{\rm att}(r) \, r^2 (1-s^2) \, \rho_1^{\prime}(z_1) \rho_1^{\prime}(z_2) \nonumber \\
&& - \frac{1}{4} \int\limits_{-\infty}^{\infty} \!\!\! dz_1 \! \int \!\! d\vec{r}_{12} \;
U_{\rm att}(r) \, r^2 (1-s^2)^2 \, z_1^2 \, \rho_0^{\prime}(z_1) \rho_0^{\prime}(z_2) \nonumber \\
&& + \frac{1}{64} \int\limits_{-\infty}^{\infty} \!\!\! dz_1 \! \int \!\! d\vec{r}_{12} \;
U_{\rm att}(r) \, r^4 (1-s^2) (1+3s^2) \, \rho_0^{\prime}(z_1) \rho_0^{\prime}(z_2) \,, \nonumber \\
\bar{k} &=& \frac{1}{2} \int\limits_{-\infty}^{\infty} \!\!\! dz_1 \! \int \!\! d\vec{r}_{12} \;
U_{\rm att}(r) \, r^2 (1-s^2) \, \rho_0^{\prime}(z_1) \, [ \, 4 \rho_{c,2}^{\prime}(z_2) - \rho_{s,2}^{\prime}(z_2) \, ] \\
&+& \frac{1}{4} \int\limits_{-\infty}^{\infty} \!\!\! dz_1 \! \int \!\! d\vec{r}_{12} \;
U_{\rm att}(r) \, r^2 (1-s^2) \, z_1^2 \, \rho_0^{\prime}(z_1) \rho_0^{\prime}(z_2) \nonumber \\
&-& \frac{1}{96} \int\limits_{-\infty}^{\infty} \!\!\! dz_1 \! \int \!\! d\vec{r}_{12} \;
U_{\rm att}(r) \, r^4 (1-s^2) (1+7s^2) \, \rho_0^{\prime}(z_1) \rho_0^{\prime}(z_2) \,. \nonumber
\end{eqnarray}
These expressions have the advantage that no reference is made to the external
field used to change the curvature. It might therefore be expected that these
expressions are independent of the way the interfacial curvature is varied.
An important disadvantage, however, is that these expressions can only be evaluated
when the second order corrections to the density profiles, $\rho_{s,2}(z)$ and
$\rho_{c,2}(z)$, are determined as well.


\begin{thebibliography}{0}
\bibitem{correctie_Blokhuis92a}
See Eq.(4.6) in ref.~\cite{Blokhuis92a}; a minus sign should be added
in the expression for $k$.
\end{thebibliography}
\end{document}